\documentclass[tighten,nolinenumbers, notrackchanges, twocolumn]{aastex631}

\usepackage{amsmath}
\usepackage{amssymb}
\usepackage{amssymb}
\usepackage{mathtools}

\usepackage{mathrsfs}
\usepackage{booktabs} 
\usepackage{morefloats}
\usepackage{longtable}
\usepackage{afterpage}

\usepackage{xcolor}
\usepackage{graphicx}
\usepackage{hyperref} 

\definecolor{mygreen}{rgb}{0,0.502,0}

\hypersetup{linkcolor=mygreen,citecolor=mygreen,urlcolor=mygreen}

\usepackage{soul}

\usepackage[makeroom]{cancel}

\graphicspath{{figs/}}

\usepackage{xspace}

\newcommand*{\XPSI}{X-PSI\xspace}
\newcommand*{\NICER}{NICER\xspace}
\newcommand*{\xmm}{XMM-Newton\xspace}
\newcommand*{\PyMultiNest}{\textsc{PyMultiNest}\xspace}
\newcommand*{\MultiNest}{\textsc{MultiNest}\xspace}

\newcommand{\msol}{$M_{\odot}$\xspace}
\newcommand{\jdbl}{PSR~J0030$+$0451\xspace}
\newcommand{\joh}{PSR~J0740$+$6620\xspace}
\newcommand*{\jof}{PSR~J0437$-$4715\xspace}
\newcommand*{\joa}{PSR~J1231$-$1411\xspace}
\newcommand*{\jos}{PSR~J0614$-$3329\xspace}

\newcommand{\xpsi}{\texttt{X-PSI}\xspace}
\newcommand{\STS}{\texttt{ST-S}\xspace}

\newcommand{\STPST}{\texttt{ST+PST}\xspace}
\newcommand{\STPDT}{\texttt{ST+PDT}\xspace}
\newcommand{\PDTU}{\texttt{PDT-U}\xspace}

\newcommand{\TT}[1]{\texttt{#1}}

\newcommand{\be}{\begin{equation}}
\newcommand{\ee}{\end{equation}}

\defcitealias{Vinciguerra:2024}{V24}

\received{}
\revised{}
\accepted{}
\published{}

\submitjournal{The Astrophysical Journal}

\shorttitle{A NICER View of \jdbl: Updated Constraints}

\shortauthors{Kini~et~al.}

\begin{document}

\title{A NICER View of \jdbl: Updated Constraints from Six Years of NICER Observations}

\email{y.kini@uva.nl}

\author[0000-0002-0428-8430]{Yves Kini}
\affiliation{Gravitation and Astroparticle Physics Amsterdam (GRAPPA), University of Amsterdam, 1098 XH Amsterdam, The Netherlands}

\author[0000-0002-3408-2759]{Lucien~Mauviard}
\affil{IRAP, CNRS, 9 avenue du Colonel Roche, BP 44346, F-31028 Toulouse Cedex 4, France}
\affil{Universit\'{e} de Toulouse, CNES, UPS-OMP, F-31028 Toulouse, France}

\author[0000-0001-6356-125X ]{Tuomo~Salmi}
\affil{Department of Physics, University of Helsinki, P.O. Box 64, FI-00014 University of Helsinki, Finland}

\author[0000-0002-1009-2354]{Anna~L.~Watts}
\affiliation{Gravitation and Astroparticle Physics Amsterdam (GRAPPA), University of Amsterdam, 1098 XH Amsterdam, The Netherlands}
\affil{Anton Pannekoek Institute for Astronomy, University of Amsterdam, Science Park 904, 1098XH Amsterdam, the Netherlands}

\author[0000-0002-6449-106X]{Sebastien~Guillot}
\affil{IRAP, CNRS, 9 avenue du Colonel Roche, BP 44346, F-31028 Toulouse Cedex 4, France}
\affil{Universit\'{e} de Toulouse, CNES, UPS-OMP, F-31028 Toulouse, France}

\author[0000-0002-9407-0733]{Bas~Dorsman}
\affil{Anton Pannekoek Institute for Astronomy, University of Amsterdam, Science Park 904, 1098XH Amsterdam, the Netherlands}

\author[0000-0002-2651-5286]{Devarshi~Choudhury}
\affil{Anton Pannekoek Institute for Astronomy, University of Amsterdam, Science Park 904, 1098XH Amsterdam, the Netherlands}

\author[0000-0001-5848-0180]{Denis~Gonz\'alez-Caniulef}
\affil{IRAP, CNRS, 9 avenue du Colonel Roche, BP 44346, F-31028 Toulouse Cedex 4, France}
\affil{Universit\'{e} de Toulouse, CNES, UPS-OMP, F-31028 Toulouse, France}

\author[0009-0005-8019-0426]{Mariska~Hoogkamer}
\affiliation{Anton Pannekoek Institute for Astronomy, University of Amsterdam, Science Park 904, 1098XH Amsterdam, the Netherlands}

\author[0000-0002-1169-7486]{Daniela~Huppenkothen}
\affiliation{Anton Pannekoek Institute for Astronomy, University of Amsterdam, Science Park 904, 1098XH Amsterdam, the Netherlands}

\author[0009-0008-3894-4783]{Christine~Kazantsev}
\affil{IRAP, CNRS, 9 avenue du Colonel Roche, BP 44346, F-31028 Toulouse Cedex 4, France}
\affil{Universit\'{e} de Toulouse, CNES, UPS-OMP, F-31028 Toulouse, France}

\author[0000-0002-0893-4073]{Matthew~Kerr}
\affil{Space Science Division, U.S. Naval Research Laboratory, Washington, DC 20375, USA}

\author[0000-0001-6573-7773]{Samaya Nissanke}
\affiliation{Gravitation and Astroparticle Physics Amsterdam (GRAPPA), University of Amsterdam, 1098 XH Amsterdam, The Netherlands}
\affiliation{Deutsches Elektronen Synchrotron DESY, Platanenallee 6, 15738 Zeuthen, Germany}
\affiliation{Deutsches Zentrum f\"ur Astrophysik (DZA), Postplatz 1, 02826 G\"orlitz, Germany}
\affiliation{Institut f{\"u}r Physik und Astronomie, Universit{\"a}t Potsdam, Haus 28, Karl-Liebknecht-Str. 24/25, 14476, Potsdam, Germany}

\author[0000-0002-5297-5278]{Paul~S.~Ray}
\affil{Space Science Division, U.S. Naval Research Laboratory, Washington, DC 20375, USA}

\author[0009-0005-7766-5638]{Pierre~Stammler}
\affil{IRAP, CNRS, 9 avenue du Colonel Roche, BP 44346, F-31028 Toulouse Cedex 4, France}
\affil{Universit\'{e} de Toulouse, CNES, UPS-OMP, F-31028 Toulouse, France}

\author[0000-0003-3068-6974]{Serena~Vinciguerra}
\affil{Independent researcher}

\begin{abstract}

Pulse-profile modeling of rotation-powered millisecond pulsars targeted by \NICER has enabled mass--radius constraints of several neutron star sources, with implications for the dense-matter equation of state. For the bright isolated pulsar PSR~J0030+0451, the inferred mass--radius was previously found to depend strongly on the assumed hot spot model. These hot-spot models yielded different mass--radius constraints, with the statistically preferred model exhibiting some mild tension with results inferred for PSR~J0437$-$4715, PSR~J0614$-$3329, and GW170817. We present an updated pulse-profile analysis of PSR~J0030+0451 using new \NICER observations obtained between 2017 July to 2023 January, increasing the number of X-ray counts by about 50\% compared to previous analyses. We jointly analyze the \NICER data with archival \xmm observations to better constrain the source spectrum and background. The new analysis significantly reduces the discrepancy between the hot spot models.  The inferred mass and radius are $M = 1.43^{+0.20}_{-0.17}\,M_\odot$ and $R_{\rm eq} = 12.68^{+1.31}_{-1.04}$~km (68\% credible intervals), reducing the tension with the results from other sources. In addition, the inferred hot spot configurations suggest the presence of intra-spot temperature gradients.

\end{abstract}


\section{Introduction}\label{sec:intro}

Rotation-powered millisecond pulsars (MSPs) are thought to form through a ``recycling" process, where an older neutron star is spun up by accreting matter from a binary companion \citep{Alpar:1982, Radhakrishnan:1982}. Studying the properties of these pulsars, particularly their masses and radii, provides a powerful means of probing one of the most outstanding questions in modern physics: the equation of state (EoS) of dense and cold matter \citep[see e.g.,][]{Lattimer:2012, Oertel:2016, Baym:2017, Tolos:2020, Yang:2019, Hebeler:2020,Chatziioannou:2025}. One of the most effective methods to infer their mass and radius is by detailed modeling of their X-ray emission using the pulse profile modeling technique, which exploits the relativistic effects associated with their high compactness and rapid rotation \citep[see, e.g.,][and references therein]{Watts:2016, Bogdanov:2019_ppm}. The X-ray emission is believed to originate from polar caps heated by the bombardment of charged particles from a magnetospheric return current \citep{Ruderman:1975, Arons:1981, Harding:2001}.

Retrieving information about the mass, radius, and surface map of a neutron star using pulse profile modeling requires high-resolution timing and spectral data. The primary goal of the Neutron Star Interior Composition Explorer \citep[NICER,][]{Gendreau:2016} is to obtain such data. Since its launch in 2017, NICER has observed several MSPs targeted for pulse profile modeling \citep[][]{Guillot:2019, Bogdanov:2019_data, Wolff:2021}. Constraints on neutron star properties have since been published for five sources: \jdbl \citep{Miller:2019, Riley:2019, Salmi:2023,Vinciguerra:2024}, \joh \citep{Miller:2021, Riley:2021, Salmi:2022, Salmi:2023, Salmi:2024,Dittmann:2024} \jof \citep{Choudhury:2024, Miller:2025}, \joa \citep{Salmi:2024_1231, Qi:2025} and \jos \citep{Mauviard:2025}.
Together, these results have provided robust constraints on the EoS of cold dense matter \cite[see, e.g.,][for recent papers]{Koehn:2024, Rutherford:2024, Golomb:2025, Huang:2025, Li:2025, Biswas:2025} and on the configuration of the X-ray emitting magnetic poles, which are inconsistent with simple centered dipole models \citep{Bilous:2019, Chen:2020, Kalapotharakos:2021, Petri:2023, Petri:2025, Cao:2026}.

While these analyses have yielded precise constraints, the inferred mass and radius can depend sensitively on the assumed complexity of the emission geometry \citep{Vinciguerra:2024}, sometimes leading to multiple statistically viable solutions with different implications for the EoS. This is particularly relevant for \jdbl, the focus of this work. Among a broad set of geometries explored by \citealt{Vinciguerra:2024} (hereafter \citetalias{Vinciguerra:2024}), two distinct hot spot models were found to yield statistically viable but different mass–radius solutions.

The first model, the \textit{single-temperature plus protruding dual-temperature} configuration (\STPDT), describes the emission using two regions: a single circular hot spot emitting uniformly at one temperature (\texttt{ST}), and a second region (\texttt{PDT}) composed of two overlapping circles with different radii and temperatures, each emitting independently (see Section~\ref{sec:emission}). The second geometry, the \textit{two protruding dual-temperature with unshared parameters} (\PDTU), is composed of two \texttt{PDT} regions, each characterized by its own set of geometric and thermal parameters. 

Using the \STPDT geometric model, \citetalias{Vinciguerra:2024} obtained a radius of $R_{\rm eq} = 11.71^{+0.88}_{-0.83}$~km and a mass of $M = 1.40^{+0.13}_{-0.12}$~\msol, whereas the \PDTU model yielded a larger radius of $R_{\rm eq} = 14.44^{+0.88}_{-1.05}$~km and a higher mass of $M = 1.70^{+0.18}_{-0.19}$~\msol. These two solutions have very different implications for the neutron-star EoS, with the former favoring softer EoS and the latter pointing to stiffer EoS. Model comparison using Bayesian evidence indicated that the \PDTU geometry is preferred by the X-ray data alone. However, the corresponding mass–radius constraints are in mild tension with results for \jof\citep{Choudhury:2024}, \jos\citep{Mauviard:2025}, and measurements from the binary neutron star merger GW170817 \citep{GW170817:eos}. In contrast, the \STPDT solution is fully consistent with these independent constraints. Furthermore, when independent information from gravitational-wave observations, radio pulsar masses, and nuclear theory is incorporated within a multi-messenger inference framework, the \STPDT model is instead favored according to Bayesian evidence \citep{Biswas:2025}.

The analysis by \citetalias{Vinciguerra:2024} built on the original work of \citet{Riley:2019}, using the same set of NICER observations obtained between 2017 July 24 and 2018 December 9 \citep[see][]{Bogdanov:2019_data}. In contrast to the earlier works \citep{Riley:2019, Miller:2019}, it used an updated \NICER instrument response and included XMM–Newton data to better constrain the \NICER spectrum/background. Since then, NICER has accumulated substantially more exposure on \jdbl, yielding roughly 50\% more X-ray counts than were included in previous analyses. The goal of this paper is therefore to update our understanding of \jdbl's properties. In particular, we aim to reassess whether distinct mass–radius constraints are still obtained when adopting the \STPDT and \PDTU hot spot models, using this new expanded data set. We also aim to assess the robustness of the inference and verify that the inferred constraints are not driven by the choice of sampling settings. We further seek to determine which emission geometry is now statistically preferred by pulse profile modeling alone, and how the resulting constraints compare both with those reported by \citetalias{Vinciguerra:2024} and with results for other 1.4 \msol MSPs (\jof\ and \jos).

The rest of this paper is organized as follows. In Section~\ref{sec:methods}, we describe the data, modeling framework, and statistical methodology. We then present the findings in Section~\ref{sec:results} and discuss their implications in Section~\ref{sec:discussion}. Finally, we summarize and conclude in Section~\ref{sec:conclusion}.

\section{Modeling}\label{sec:methods}

Similar to most mass–radius analyses obtained using pulse profile modeling of NICER data (see e.g. \citealt{Miller:2019, Riley:2019, Miller:2021, Riley:2021,
Choudhury:2024, Dittmann:2024, Salmi:2024}, \citetalias{Vinciguerra:2024}, \citealt{Mauviard:2025, Miller:2025}), we adopt the theoretical framework described in \citet{Bogdanov:2019_ppm} (see also references therein). We summarize, in Section~\ref{sec:modeling},   the key components and assumptions of this theoretical framework used to generate model X-ray pulse profiles for a given set of source properties, and describe the statistical methodology used to explore the parameter space. We also describe, in Section~\ref{sec:data}, the observational data used in this analysis, together with the reduction procedure. 

Throughout this work, we use the publicly available X-ray Pulse Simulation and Inference package \citep[\texttt{X-PSI}\footnote{\href{https://github.com/xpsi-group/xpsi}{https://github.com/xpsi-group/xpsi}}, v.3.1.0;][]{Riley:2023} to model X-ray pulse profiles and constrain the properties of \jdbl. For this source, using \texttt{X-PSI} high-resolution or low-resolution settings as described in detail in Section 2.3.1 of \citet{Vinciguerra:2023} does not lead to significant changes in the inferred results \citepalias[see][]{Vinciguerra:2024}. These settings specify, among other aspects, the energy grid used to compute the initial signals, the discretization of the emitting hot spot surface, and the phase resolution in the source frame. We therefore adopt the \texttt{X-PSI} low-resolution configuration, using the parameter values listed in Section 2.3.1 of \citet{Vinciguerra:2023}, since this choice provides a good balance between computational efficiency and inference accuracy.

\subsection{X-Ray Event Data}\label{sec:data}

The new NICER dataset includes all exposures from 2017 July 24 to 2023 January 13, all being prior to the NICER light-leak that occurred in May 2023. Observations during the light-leak are characterized by a high background, and even if they can be corrected\footnote{\url{https://heasarc.gsfc.nasa.gov/docs/nicer/analysis_threads/light-leak-analysis/}}, they account for only $2.4\%$ of the total exposure, hence we decided to not use this data. 

We used the \texttt{nicerl2} task from the \textsc{Heasoft} v6.33 software \citep{Heasoft:2014} along with the \texttt{xti20240206} calibration file to preprocess the data using the default parameters. Using the \texttt{psrpipe} task from NICERsoft\footnote{\href{https://github.com/paulray/NICERsoft}{https://github.com/paulray/NICERsoft}}, we further reduced the data in a similar manner as in \citet{Riley:2019}, with a few noticeable improvements. First, the few detectors considered ``hot" among the 52 onboard are no longer consistently removed, but now detectors that exhibit abnormally high count rates ($3\sigma\,$ above the median of all detectors) are individually excluded. Second, we impose the median undershoot rate among detectors to be lower than $200\,\rm{cts\,s}^{-1}$ to avoid high optical photon flux (which would cause increased background at energies under $0.35$~keV). Finally, we exclude time intervals where the count rate in the $2.0-10.0\,$keV energy band exceeded $1.5 \, \rm{cts\,s}^{-1}$. This energy band was chosen to be a good proxy for background, as the contribution of the pulsar in this band is negligible. The ancillary response file (ARF) and the response matrix files (RMF) were subsequently extracted with the \texttt{nicerl3-spect} task, with the  ARF being scaled by the average number of detectors available.

\begin{figure}
    \centering
     \includegraphics[width=\columnwidth]{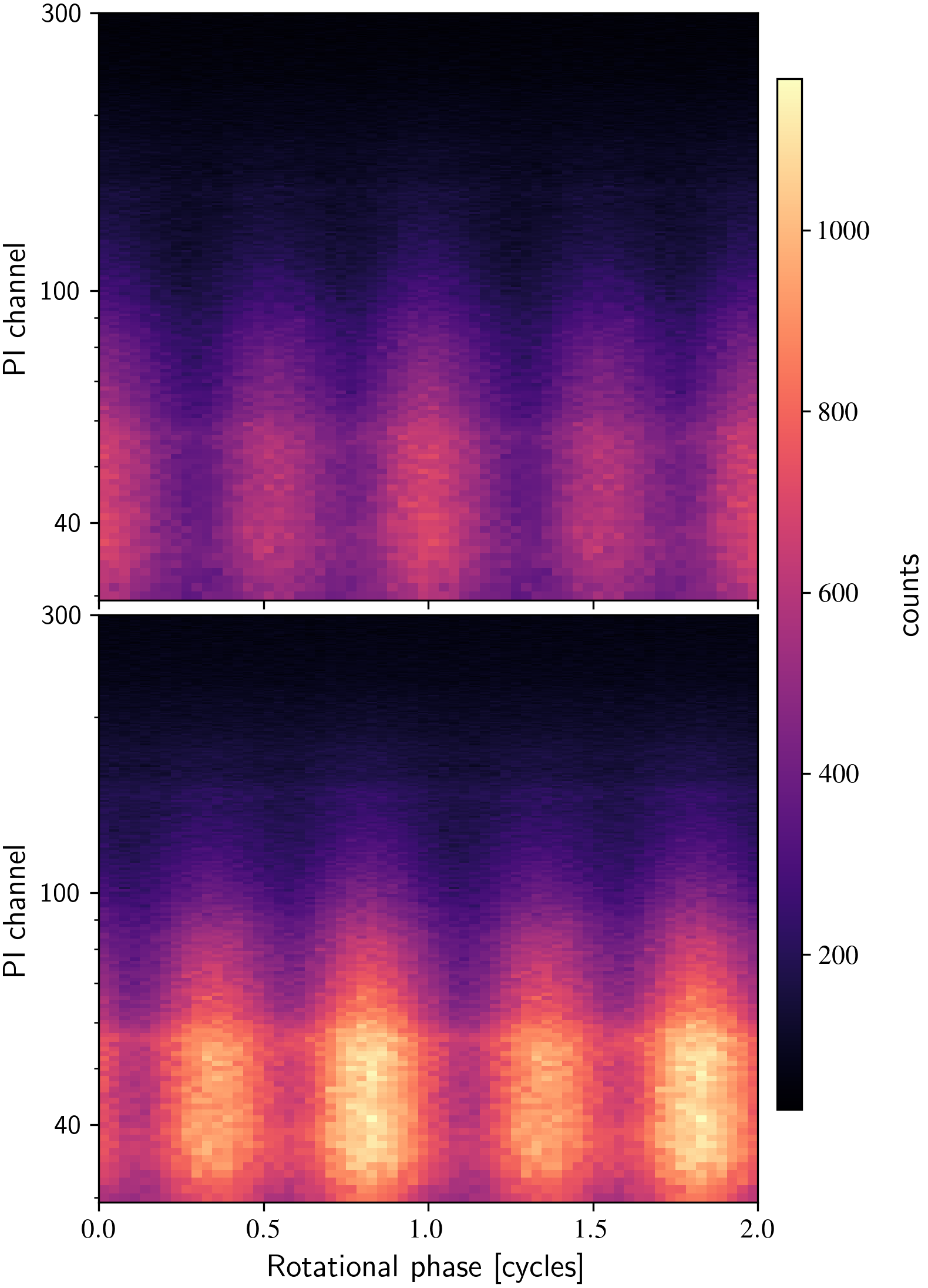}
      \caption{NICER phase--channel--resolved counts for \jdbl. Counts are folded on the rotational phase and shown over two cycles for clarity; the second cycle repeats the first. The top panel shows the dataset analyzed by \citetalias{Vinciguerra:2024}, while the bottom panel shows the new data used in this work. Both analyses use the PI channels subset $[30,300)$, corresponding nominally to the photon energy range $[0.3$--$3.0)$~keV. The rotational phase is divided into 32 equally spaced bins per cycle. The horizontal features present around PI channels 35 and 55 are due to fluorescence lines originating in the Earth’s atmosphere \citep[see][for a detailed discussion]{Mauviard:2025}.}
      \label{fig:lightcures}
\end{figure}

This processing results in a final exposure time of 2,974 ks, out of the initial 4,029 ks available, which represents a 53\% increase compared to the data used by \citet{Riley:2019} and \citetalias{Vinciguerra:2024}. The remaining events were phase-folded using the \texttt{photonphase} task from the PINT software \citep{Luo:2012_pint} along with a radio timing solution spanning the entire duration of the NICER data. For our analysis, we used the events from the Pulse Invariant\footnote{Here “pulse” refers to the detector charge pulse height used to determine photon energy, not to the neutron star rotational pulse profile.} (PI) channels [30,300), corresponding to the 0.3$-$3.0 keV energy band. The final resulting data and the comparison to the previous dataset from \citet{Riley:2019} used by \citetalias{Vinciguerra:2024} is shown in Figure \ref{fig:lightcures}.

As no new \xmm observation has been made since the last analysis, we used the same data as in \citetalias{Vinciguerra:2024} and provided in \citet{Vinciguerra_Zenodo:2024}.

\subsection{Pulse Profile Modeling Using X-PSI}\label{sec:modeling}

As described in \citet{Bogdanov:2019_ppm} and references therein, the X-rays observed by NICER are assumed to originate from hot spots on the surface of the neutron star (see Section~\ref{sec:emission} for a description of the hot spot models). The star is modeled as an oblate spheroid to account for rotational flattening, with the external spacetime described by the Schwarzschild metric \citep[see e.g,][]{Cadeau:2007, Morsink:2007}. Rotational effects such as relativistic Doppler boosting and aberration are then applied after computing the light bending. The stellar oblateness and the effective gravitational acceleration at each surface location are computed using fitting formulae calibrated on relativistic stellar-structure models based on a representative set of EoS \citep[see][for more details]{AlGendy:2014}. 

In the modeling, the emitted X-rays from the surface hot spots are assumed to propagate through and interact with a stellar atmosphere, which modifies both their spectral shape and angular emission pattern. The properties of the emergent radiation are therefore sensitive to the assumed atmospheric composition. Following  \citet{Riley:2019} and \citetalias{Vinciguerra:2024}, we assume for \jdbl\ a fully ionized, non-magnetic hydrogen atmosphere and use the NSX atmosphere models\footnote{See \citet{Salmi:2023} and \citet{Vinciguerra:2023} for details of the atmosphere implementation in \texttt{X-PSI}.} \citep{Ho:2001,Ho:2009}.

In addition to atmospheric effects, interstellar absorption along the line of sight is included in the model. We parameterize this attenuation through the hydrogen column density $N_{\rm H}$ using the \texttt{tbnew} model \citep{Wilms:2000}.

To compare the model predictions with the data, the theoretical signal is folded through the instrumental response of each detector, producing either a NICER pulse profile or an \xmm\ spectrum. The response matrices account for the effective area and energy redistribution of the instruments. To incorporate cross-calibration uncertainties between NICER, MOS1, and MOS2, we introduce an energy-independent scaling factor, $\alpha$, that multiplies each ARF (see Section~2.2 of \citealt{Salmi:2022} for a detailed discussion). Following previous analyses (see e.g., \citetalias{Vinciguerra:2024}; \citealt{Salmi:2024_1231}), we assume a common scaling factor for MOS1 and MOS2.

\subsubsection{Emission regions models}\label{sec:emission}
As in previous \NICER data analyses using \xpsi, we assume that the observed X-ray pulses originate from two distinct hot regions. This is mainly motivated by previous theoretical works \citep[see, e.g.,][]{Harding:2001,Harding:2011, Timokhin:2013,Kalapotharakos:2014, Gralla:2017,Lockhart:2019, Kalapotharakos:2021}. Each hot region is parametrized using one or two circular shapes, enabling various emitting geometries. 
The resulting hierarchy of models spans from the simplest \STS\ (\textit{Single-temperature Regions with Antipodal Symmetry}) configuration to the most complex \PDTU geometry \citep[see Figure 1 of][]{Vinciguerra:2023}. These models are nested such that higher-complexity configurations can, in principle, approximate those with fewer degrees of freedom \citep[see][for a detailed description of all the models]{Riley:2019}. In this sense, the \PDTU\ model can approximately mimic all lower-complexity geometries. Among the available geometries, we focus in this work on the two configurations that were found by \citetalias{Vinciguerra:2024} to best reproduce the NICER data: the \STPDT and \PDTU models.\footnote{Neither the \STPDT\ nor the \PDTU\ models were explored in \citet{Riley:2019}. However, the mass--radius constraints obtained with the \STPDT\ model in \citetalias{Vinciguerra:2024} are consistent with those derived using the \STPST\ model in \citet{Riley:2019}.}

In the \STPDT model, the X-ray–emitting surface is described by two distinct regions: a circular single-temperature region (\texttt{ST}) and a protruding dual-temperature region (\texttt{PDT}). The \texttt{ST} component is characterized by a uniform temperature across its entire area. The \texttt{PDT} component consists of two circular sub-regions with unique uniform temperatures and radii. These two circles are required to be contiguous, that is, one may be fully or partially embedded within the other, or, at minimum, the two must be in contact. In the overlapping region, the emission is only accounted for from one circle (the \textit{superseding} region), while the contribution from the other circle (the \textit{ceding} region) is included only in its non-overlapping area. Such a parametrization helps form complex shapes. In this configuration, the \texttt{ST} region is referred to as the \textit{primary} (\textit{superseding}) hot spot, while the \texttt{PDT} region is designated as the \textit{secondary} hot spot \citep[see][for details on the nomenclature and geometry]{Riley:2019, Vinciguerra:2023}.

The second configuration, the \PDTU hot spot model, features two independent \texttt{PDT} regions with unshared parameters. This means that each \texttt{PDT} region has its own independent set of temperatures and angular radii. The \PDTU geometry is more flexible than the \STPDT configuration and can exactly reproduce it as a special case. In previous works (see e.g., \citetalias{Vinciguerra:2024}; \citealt{Salmi:2024_1231}), the spot with the smaller colatitude is typically defined as the \textit{primary}, with the other labeled as the \textit{secondary}. Here, we instead define the spot with the smaller colatitude as the \textit{secondary}, and label the other as the \textit{primary}, to facilitate comparison with the \STPDT\ model.

With the above assumptions about relativistic ray tracing, interstellar absorption, and instrumental responses, the model is fully specified by a set of parameters. These parameters are summarized in Table~\ref{tab:model_parameters}, together with their corresponding priors.

\begin{deluxetable*}{lll}
\tablewidth{0pt}
\tabletypesize{\scriptsize}
\tablecaption{Model parameters and adopted priors. Shared parameters are common to both the \STPDT\ and \PDTU\ models.\label{tab:model_parameters}}
\tablehead{\colhead{Parameter} & \colhead{Description} & \colhead{Prior PDF (density and support)}}
\startdata
\multicolumn{3}{l}{\textit{Shared parameters (\STPDT\ and \PDTU)}} \\
$M$ [$\mathrm{M_\odot}$] & Gravitational mass & $M\sim\mathcal{U}(1.0,3.0)$ \\
$R_{\rm eq}$ [km] & Equatorial circumferential radius & $R_{\rm eq}\sim\mathcal{U}(3r_g(1.0),16.0)$\tablenotemark{a} \\
& & Compactness: $R_{\rm polar}/r_g(M)>3$; surface gravity: $13.7\leq \log_{10}g(\theta)\leq15.0$,~$\forall\theta$ \\
$D$ [kpc] & Distance to the source & $D\sim\mathcal{N}(0.325,0.009^2)$ \\
$\cos i$ & Cosine of the inclination angle to the spin axis & $\cos(i)\sim\mathcal{U}(0.0,1.0)$ \\
$N_{\rm H}$ [$10^{20}$cm$^{-2}$] & Interstellar absorption column & $N_{\rm H}\sim\mathcal{U}(0.0,5.0)$ \\
$\alpha_{\rm XTI}$ & NICER effective-area scale factor & $\alpha_{\rm XTI}\sim\mathcal{N}(1.0,0.104^2)$\tablenotemark{b} \\
$\alpha_{\rm MOS1}$ & XMM--Newton MOS1 effective-area scale factor &
$\alpha_{\rm MOS1}\mid\alpha_{\rm XTI} \sim
\mathcal{N}\!\big(1+\rho(\alpha_{\rm XTI}-1),\,0.104^2(1-\rho^2)\big)$\tablenotemark{c} \\
$\theta_{\rm p,s}$ [radian] & Colatitude of the primary superseding spot & $\cos(\theta_{\rm p,s})\sim\mathcal{U}(-1.0,1.0)$ \\
$\zeta_{\rm p,s}$ [radian] & Angular radius of the primary superseding spot & $\zeta_{\rm p,s}\sim(0.0,\pi/2)$ \\
$\log_{10}(T_{\rm p,s}[\rm K])$ & Effective temperature of the primary superseding spot & $\log_{10}(T_{\rm p,s}[\rm K])\sim\mathcal{U}(5.1,6.8)$ \\
$\phi_{\rm p}$ [cycle] & Initial rotational phase of the primary superseding spot & $\phi_{\rm p}\sim\mathcal{U}(-0.25,0.75)$ \\
$\theta_{\rm s,s}$ [radian] & Colatitude of the secondary superseding spot & $\cos(\theta_{\rm s,s})\sim\mathcal{U}(-1.0,1.0)$ \\
$\zeta_{\rm s,s}$ [radian] & Angular radius of the secondary superseding spot & $\zeta_{\rm s,s}\sim(0.0,\pi/2)$ \\
$\log_{10}(T_{\rm s,s}[\rm K])$ & Effective temperature of the secondary superseding spot & $\log_{10}(T_{\rm s,s}[\rm K])\sim\mathcal{U}(5.1,6.8)$ \\
$\phi_{\rm s}$ [cycle] & Initial rotational phase of the secondary superseding spot & $\phi_{\rm s}\sim\mathcal{U}(-0.25,0.75)$ \\
$\theta_{\rm s,c}$ [radian] & Colatitude of the secondary ceding spot & $\cos(\theta_{\rm s,c})\sim\mathcal{U}(-1.0,1.0)$ \\
$\zeta_{\rm s,c}$ [radian] & Angular radius of the secondary ceding spot & $\zeta_{\rm s,c}\sim(0.0,\pi/2)$ \\
$\log_{10}(T_{\rm s,c}[\rm K])$ & Effective temperature of the secondary ceding spot & $\log_{10}(T_{\rm s,c}[\rm K])\sim\mathcal{U}(5.1,6.8)$ \\
$\chi_{\rm s,c}$ [radian] & Azimuthal offset of the secondary ceding spot & $\chi_{\rm s,c}\sim\mathcal{U}(-\pi,\pi)$ \\ \\ \hline
\multicolumn{3}{l}{\textit{\PDTU-specific parameters}} \\ 
$\theta_{\rm p,c}$ [radian] & Colatitude of the primary ceding spot & $\cos(\theta_{\rm p,c})\sim\mathcal{U}(-1.0,1.0)$ \\
$\zeta_{\rm p,c}$ [radian] & Angular radius of the primary ceding spot & $\zeta_{\rm p,c}\sim(0.0,\pi/2)$ \\
$\log_{10}(T_{\rm p,c}[\rm K])$ & Effective temperature of the primary ceding spot & $\log_{10}(T_{\rm p,c}[\rm K])\sim\mathcal{U}(5.1,6.8)$ \\
$\chi_{\rm p,c}$ [radian] & Azimuthal offset of the primary ceding spot & $\chi_{\rm p,c}\sim\mathcal{U}(-\pi,\pi)$ \\
\enddata
\tablenotetext{a}{$r_g$ is the Solar Schwarzschild gravitational radius.}
\tablenotetext{b}{We assume a 10\% uncertainty in the shared calibration factor and a 3\% uncertainty in the telescope-specific factor, which combine in quadrature to give a total uncertainty of 10.4\% (see Section~2.2 of \citealt{Salmi:2022} for a detailed discussion).}
\tablenotetext{c}{The NICER and XMM--Newton effective-area scale factors are treated as correlated to account for shared calibration uncertainties; the conditional form is used to implement this correlation with $\rho=0.916$.}
\end{deluxetable*}

\subsubsection{Sampling and convergence}\label{sec:sampling}

To explore the parameter space, we employ the nested sampling algorithm \texttt{MultiNest} \citep{MultiNest_2008, MultiNest_2009, MultiNest_2019, pymultinest:2014} as in \citetalias{Vinciguerra:2024}. For the joint \NICER and \xmm analysis with the \STPDT and \PDTU models, \citetalias{Vinciguerra:2024} performed the parameter space exploration using 1000 live points, a sampling efficiency of 0.8, and with multi-mode disabled. In this work, our headline results for both the \STPDT and \PDTU models are obtained using improved 10{,}000 \texttt{MultiNest} live points, a sampling efficiency of 0.3, and with multi-mode also disabled. We also carried out a set of exploratory runs in which we varied the number of live points, the sampling efficiency, and with multi-modal mode enabled or disabled (see Fig.~\ref{fig:mass_radius_explore} in Appendix~\ref{appendix:runs}). Beyond a certain threshold in the sampling settings, further increases in the number of live points or reductions in the sampling efficiency do not lead to any significant change in the inferred posterior distributions. We note that there is no method that can prove with absolute certainty the convergence of a sampling algorithm in practical applications. Nevertheless, the consistency of the results across a wide range of sampler settings \citep[see][for practical advice]{Higson:2019} gives us confidence in the robustness of our results within the quoted uncertainties (see Section~\ref{sec:convergence} for further discussion of convergence).

\section{Results}\label{sec:results}

In what follows, we report the results of the joint \NICER and \xmm analysis. All quoted uncertainties correspond to the 68\% credible intervals. In Section~\ref{sec:M_R}, we present the resulting constraints on the mass and radius of \jdbl\ obtained with both hot spot models. The corresponding hot spot geometries are presented in Section~\ref{sec:geometry}. The joint posterior distributions for all model parameters, together with the full posterior samples, are available on Zenodo\footnote{\url{https://doi.org/10.5281/zenodo.18741942}}.

Using joint \NICER and \xmm observations, the modeling assumptions described in Section~\ref{sec:methods}, and Bayesian inference with \texttt{MultiNest}, we analyze the data and find that both hot spot configurations, \STPDT\ and \PDTU, provide a good description of both the new \NICER and \xmm data. Inspection of the residuals (Figure~\ref{fig:residuals} in Appendix~\ref{appendix:residuals}) shows no systematic trends or clustering in either case. Moreover, the residual distributions are visually similar for the two models, suggesting comparable fit quality. Nevertheless, a comparison of the Bayesian evidences (still) shows a clear preference for the \PDTU\ model, which yields a significantly higher evidence ($\log Z_{\mathrm{\texttt{PDT-U}}} = -44475.77\pm 0.08$) than the \STPDT\ model ($\log Z_{\mathrm{\texttt{ST+PDT}}} = -44483.68\pm 0.07$).

\subsection{Mass and radius}\label{sec:M_R}

\begin{figure*}
    \centering
     \includegraphics[width=0.75\linewidth]{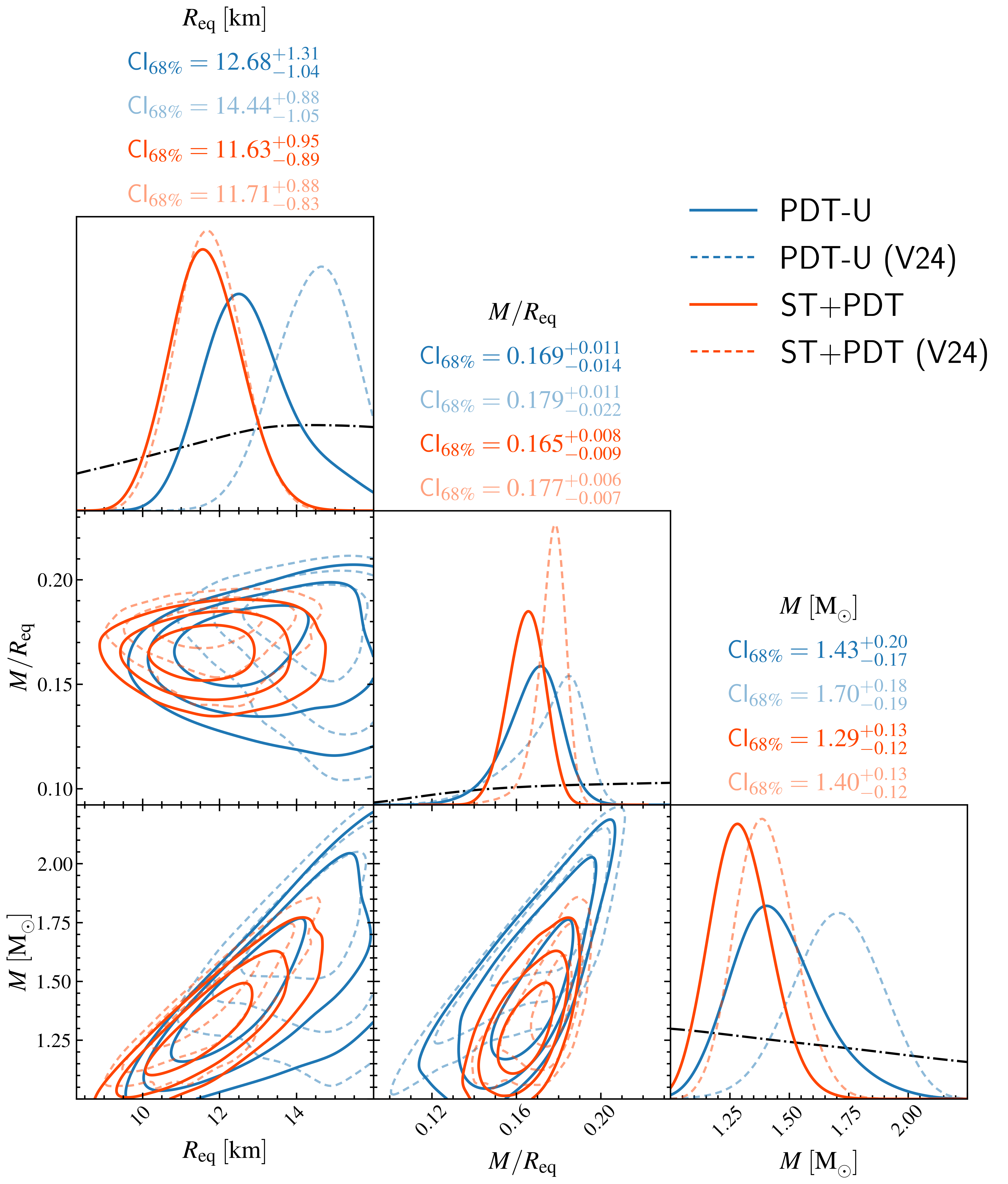}
       \caption{Posterior distributions of the equatorial radius $R_{\rm eq}$, compactness $M/R_{\rm eq}$, and mass $M$. Solid curves show the results obtained in this work, while dashed curves correspond to the analysis of \citetalias{Vinciguerra:2024}. Blue curves denote the \PDTU\ model and orange curves the \STPDT\ model. The two-dimensional panels show the joint marginal posterior distributions, with contours enclosing the 68\%, 95\%, and 99\% credible regions. The diagonal panels show the marginalized one-dimensional posterior distributions for each parameter and the black dotted curves indicate the corresponding one-dimensional priors. For each model and analysis, the median and 68\% credible intervals are indicated in the corresponding color.}
      \label{fig:mass_radius}
\end{figure*}

The mass and radius posterior distributions are shown in Figure~\ref{fig:mass_radius} for the new data set (solid lines). These results are compared to those obtained by \citetalias{Vinciguerra:2024} (dashed lines). For the \STPDT\ model, the radius posterior distribution inferred from the new NICER data is almost identical to that of \citetalias{Vinciguerra:2024}, with an inferred radius of $R_{\rm eq} = 11.63^{+0.95}_{-0.89}$ km. The newly inferred mass, $M= 1.29^{+0.13}_{-0.12}$ \msol, however, has shifted slightly toward lower values, while remaining broadly consistent with the findings of \citetalias{Vinciguerra:2024}.

A more significant change is observed for the \PDTU\ model. Regions of larger radii that were favored by the previous data are now disfavored with the new NICER observations and improved sampler settings. This leads to a decrease in the median  inferred radius by about 1.8 km, compared to the value reported in \citetalias{Vinciguerra:2024} ($R_{\rm eq}= 14.44^{+0.88}_{-1.05}$ km) to $R_{\rm eq} = 12.68^{+1.31}_{-1.04}$ km. A similar trend is seen for the mass, which shifts from $M = 1.70^{+0.18}_{-0.19}$ \msol in \citetalias{Vinciguerra:2024} to $M = 1.43^{+0.20}_{-0.17}$ \msol.

Interestingly, the width of the 68\% credible interval for the mass remains unchanged relative to \citetalias{Vinciguerra:2024} for both hotspot models ($\Delta M = 0.37$~\msol\ for \PDTU\ and $\Delta M = 0.25$~\msol\ for \STPDT). In contrast, the radius uncertainty increases by approximately 22\% and 8\% for the \PDTU\ and \STPDT models, respectively. This broadening is driven primarily by our more conservative nested-sampling settings\footnote{With identical sampler settings, the new data provide substantially tighter constraints than \citetalias{Vinciguerra:2024} (see Figure~\ref{fig:mass_radius_explore})}. We verified, however, that beyond a certain point increasing the number of live points or reducing the sampling efficiency further has a negligible impact on the posterior widths (see Appendix~\ref{appendix:runs}).

Overall, the mass and radius inferred from both hot spot models are now more consistent. The same trend is also seen in the hot spot geometry, which we report next.

\subsection{Hot spot geometry}\label{sec:geometry}

The inferred hot spot geometries for the two emission models are very similar, as shown in Figure~\ref{fig:hotspots_posteriors} (see also Figure~\ref{fig:surf_maps} of Appendix~\ref{appendix:geometry}). In both cases, the primary emitting region is located in the southern hemisphere, with a median colatitude of $\Theta_{p,s}\simeq2.3$--$2.4$~rad ($\Theta_{p,s}=2.3^{+0.1}_{-0.2}$~rad for \PDTU\ and $\Theta_{p,s}=2.4^{+0.1}_{-0.1}$~rad for \STPDT). The corresponding angular radii are also consistent between the two models, with median values of $\zeta_{p,s}=0.08^{+0.01}_{-0.02}$~rad (\PDTU) and $\zeta_{p,s}=0.11^{+0.01}_{-0.01}$~rad (\STPDT). Likewise, the inferred temperatures of these primary regions are nearly identical, with the median $\log_{10}(T_{p,s}\,[\mathrm{K}])\simeq6.1$ in both models. However a notable difference arises in the structure of the primary region in the \PDTU\ model, which includes a second hot sub-region that has no counterpart in the \STPDT\ configuration. This additional component yields a relatively extended but cool emitting area, with $\zeta_{p,c}=0.59^{+0.60}_{-0.22}$~rad and $\log_{10}(T_{p,c}\,[\mathrm{K}])=5.67^{+0.12}_{-0.24}$.

The secondary components of the \STPDT\ and \PDTU\ configurations also exhibit very similar properties. In both cases, they are located in the northern hemisphere, with median colatitudes around $\Theta\simeq1.3$--$1.4$~rad. The cooler component (the \emph{superseding} region) of each PDT sub-region has similar median temperatures, with $\log_{10}(T_{s,s}\,[\mathrm{K}])=5.89^{+0.03}_{-0.04}$ for \PDTU\ and $5.99^{+0.01}_{-0.01}$ for \STPDT, and comparable angular radii of order $\zeta_{s,s}\sim0.1$~rad. The hotter component (the \emph{ceding} region) of the PDT sub-regions also yields nearly identical results in both models: an extremely small emitting area, with $\zeta\sim0.01$~rad, and a very high temperature, $\log_{10}(T\,[\mathrm{K}])\simeq 6.2-6.4$. The two models also yield consistent phase offsets ($\phi_{p}, \phi_{s}$) between their respective emitting components (\texttt{ST} and \texttt{PDT} in \STPDT, and the two \texttt{PDT} regions in \PDTU).

Overall, the posterior distributions of most geometric parameters in the \PDTU\ model appear mildly bimodal, with one of the modes closely matching the parameter values favored by the \STPDT\ configuration. The corresponding posteriors obtained with the \STPDT\ model are generally much narrower. This suggests that the \STPDT\ solution can be interpreted as a restricted subset of the broader \PDTU\ parameter space.

\begin{figure*}
    \centering
     \includegraphics[width=0.98\linewidth]{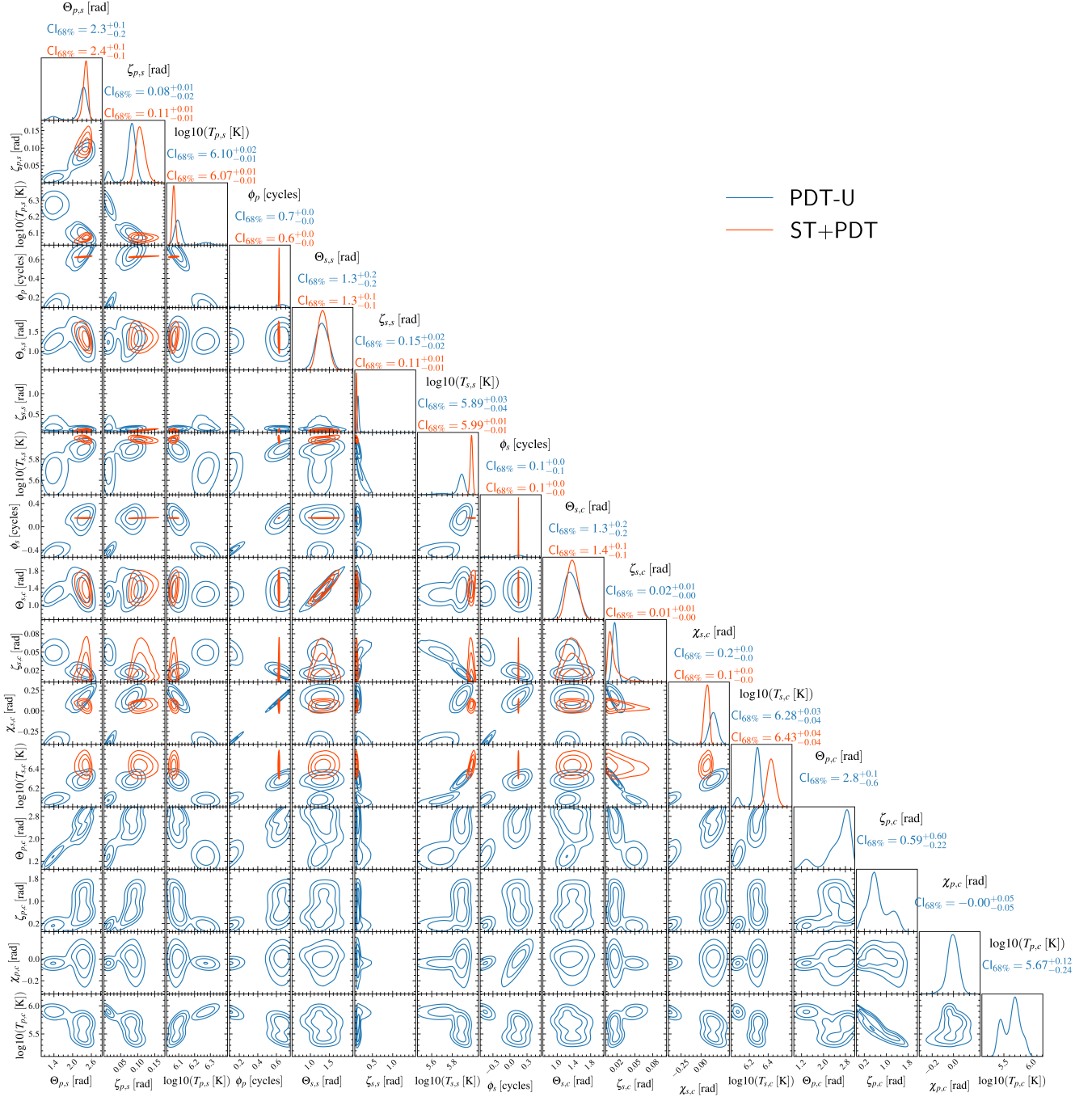}
      \caption{Posterior distributions of hot spots parameters for \jdbl. Blue curves denote the \PDTU\ model and orange curves the \STPDT\ model. The two-dimensional panels show the joint marginal posterior distributions, with contours enclosing the 68\%, 95\%, and 99\% credible regions. The diagonal panels show the marginalized one-dimensional posterior distributions for each parameter. For each model and analysis, the median and 68\% credible intervals are indicated in the corresponding color.}
      \label{fig:hotspots_posteriors}
\end{figure*}

\section{Discussion}\label{sec:discussion}

Using new NICER observations of \jdbl in combination with XMM-Newton data (the same XMM-Newton data as in \citetalias{Vinciguerra:2024}), we present an updated analysis of its mass--radius constraints and surface emission properties. Our revised modeling yields a mass of $M= 1.29^{+0.13}_{-0.12}$ \msol and a radius of $ R_{\rm eq} = 11.63^{+0.95}_{-0.89}$~km for the \STPDT\ model, and a mass of $M= 1.43^{+0.20}_{-0.17}$ \msol with a radius of $R_{\rm eq} = 12.68^{+1.31}_{-1.04}$ km for the \PDTU\ model.

\subsection{Mass and radius}\label{sec:M_R_discussion}

\begin{figure}
    \centering
     \includegraphics[width=\columnwidth]{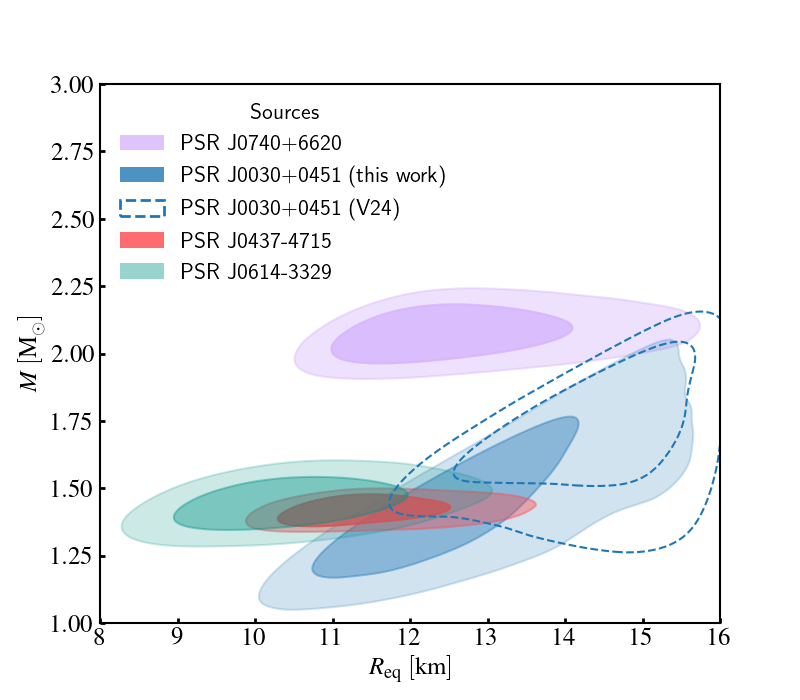}
      \caption{Comparison of mass-radius constraints from \NICER and \xmm observations. The filled contours show the 68\% and 95\% credible regions for \jdbl inferred in this work (\PDTU model), while dashed contours indicate the previous constraints from \citetalias{Vinciguerra:2024}. Also shown are published constraints for \joh, \jof, and \jos. The updated \PDTU solution shifts toward lower mass and radius, reducing the previously noted tension with other $\sim 1.4$~\msol RMPs.
      }
      \label{fig:mr_compare}
\end{figure}

A comparison of the Bayesian evidences shows that the new dataset still favours the \PDTU\ model. At the same time, the updated results for \PDTU\ give both a different spot geometry and a lower radius, consistent with a softer EoS than inferred in \citetalias{Vinciguerra:2024}. This model yields a substantial shift, with the median radius decreasing by about $1.8$~km and the median mass by approximately $0.3$~\msol. As a result, the \PDTU\ mass--radius constraints move significantly closer to those obtained with \STPDT\ and to those inferred for other $\sim 1.4$~\msol RMPs (see Figure~\ref{fig:mr_compare}). This reduces the mild tension previously noted between the \PDTU\ solution of \citetalias{Vinciguerra:2024} and the results for \jof, \jos, and GW170817. In contrast, the impact of the new \NICER\ observations on the \STPDT\ model's mass--radius constraints is modest, with the inferred mass and radius remaining largely consistent with the analysis of \citetalias{Vinciguerra:2024}.

At first sight, the magnitude of the shift observed for \PDTU\ may appear surprising, given that the new dataset contains roughly $50\%$ more counts than that used in \citetalias{Vinciguerra:2024}. Since statistical precision typically improves as $\sqrt{N}$, such an increase would naively be expected to reduce credible interval widths by only $\sim20\%$, rather than to move the dominant posterior mode by nearly $2$~km in radius and $\sim0.3$~\msol\ in mass. The fact that the \STPDT\ solution remains stable while the \PDTU\ posterior shifts significantly indicates that the difference is not driven purely by improved counting statistics.

The likelihood surface for the \PDTU\ model appears to be mildly bimodal. One mode corresponds to a solution with relatively high mass ($M\sim1.7$~\msol) and large radius ($R_{\rm eq}\sim14$~km), which is the configuration found in \citetalias{Vinciguerra:2024} using the earlier \NICER\ observations. The second mode is characterized by a lower mass and smaller radius, with $M\sim1.3$~\msol\ and $R_{\rm eq}\sim11$~km, and coincides with the solution obtained with the \STPDT\ model both in this analysis and in \citetalias{Vinciguerra:2024}. 

With the more conservative nested-sampling configuration adopted here, the relative weighting of these two modes changes: the high-mass, large-radius solution is significantly downweighted, while the lower-mass solution becomes dominant, although the former is not completely eliminated. The new data further disfavour parts of the previously preferred parameter region, reinforcing this shift. As a consequence, the posterior distribution retains support in two separated regions of parameter space, naturally leading to broader marginal posteriors in the \PDTU\ case.

\subsection{Hot spot geometry}\label{sec:geometry_discussion}

The new dataset also leads to a change in the inferred surface geometry for the \PDTU model. In \citetalias{Vinciguerra:2024}, configurations in which both hot spots were located in the same hemisphere or in opposite hemispheres were found to be equally likely. With the updated NICER observations, however, configurations in which both spots lie in the same hemisphere are now strongly disfavoured (see Figure~\ref{fig:surf_maps}). Despite this shift, the preferred geometries still do not exactly correspond to a simple centered dipole. 

In the \PDTU case, the posterior distribution of the spot parameters is also bimodal, with one of the modes closely matching the geometry found by the \STPDT model (see Figure~\ref{fig:hotspots_posteriors}). Both hot spot models yield emitting regions with very similar properties in the northern hemisphere: a moderately extended spot of angular radius $\sim0.1$~rad emitting at $\log_{10}(T\,[\mathrm{K}])\simeq6.1$, together with a much smaller region (angular radius $\sim0.01$~rad) with a significantly higher temperature ($\log_{10}(T\,[\mathrm{K}])\simeq 6.2-6.4$). This configuration -- a small, hot component embedded within a larger, cooler region, similar to that inferred for \jof \citep{Choudhury:2024} -- may point to the presence of a temperature gradient across the hot spot, which is not unexpected in realistic surface emission scenarios \citep[see e.g.][]{Harding:2001, Harding:2011,Gralla:2017,Lockhart:2019}. In the \STPDT configuration, the emitting region in the southern hemisphere has almost identical properties (both in temperature and angular size) to the larger northern spot. In the \PDTU model, the southern hemisphere also hosts an emitting region with properties comparable to the larger northern spot, but this hot component is embedded within a much cooler structure, with $\log_{10}(T\,[\mathrm{K}])\simeq5.5$. This again suggests the presence of a temperature gradient. The cooler component can extend over a substantial fraction of the stellar surface, in some cases covering up to nearly half of the star, which may indicate that a large portion—or possibly the entire—surface is sufficiently hot to contribute detectable X-ray emission. If the entire surface is indeed emitting (at lower temperatures), a significant fraction of this emission could fall outside the \NICER band, potentially in the ultraviolet. Future Hubble Space Telescope observations might therefore help to better constrain the surface emission properties, and consequently the neutron star mass and radius.

\subsection{Sampling robustness}\label{sec:convergence}

In this analysis, we performed parameter inference using \texttt{MultiNest}, as it provides Bayesian evidence estimates suitable for model comparison, and is well suited to exploring likelihoods with multimodal structure \citep[see e.g.,][]{MultiNest_2008, MultiNest_2009, MultiNest_2019}, a common feature of pulse profile modeling. In addition, \texttt{MultiNest} has been successfully applied across a wide range of scientific applications, from particle physics to cosmology \citep[see e.g., ][]{Ashton:2022,Albert:2025}

In the specific case of pulse profile modeling, we have conducted several injection studies that showed the robustness and reliability of \texttt{MultiNest}, given suitable sampler settings, when applied to this problem \citep{Kini:2023,Kini:2024a,Vinciguerra:2023,Salmi:2024, Hoogkamer:2025}. In addition, \citet{Hoogkamer:2025} performed a cross-comparison study using \texttt{UltraNest}---which is known to provide robust results but incurring higher computational expense---and obtained identical results to those from \texttt{MultiNest}, with the same credible intervals and medians.

For the new dataset of \jdbl, we performed a series of convergence tests for both hot spot models (see Appendix~\ref{appendix:runs}). These tests explored variations in the sampling efficiency, the number of live points, and -- for the \STPDT\ model -- the use of multimodal sampling. While such tests cannot guarantee strict convergence, they can provide a practical assessment of the robustness of the inferred posterior distributions \citep{Higson:2019}. We find that beyond certain parameter choices, variations in the sampling setup have only a minor impact on the results. For example, in the \PDTU case (see Figure~\ref{fig:mass_radius_explore} of Appendix~\ref{appendix:runs} for \STPDT), increasing the number of live points from 4{,}000 to 10{,}000 (at a sampling efficiency of $0.3$) only marginally broadens the 95\% and 99\% credible regions. Similarly, reducing the sampling efficiency from $0.8$ to $0.3$ (with 4{,}000 live points) has almost no impact on the inferred constraints (see Figure~\ref{fig:mass_radius_explore}). By contrast, the computational cost increases substantially: from approximately 300{,}000 to over 4{,}800{,}000 core-hours when increasing the number of live points, and from about 100{,}000 to 300{,}000 core-hours when reducing the sampling efficiency. Such modest changes in the credible intervals are unlikely to have any meaningful impact on the inferred equation-of-state parameters, while incurring a disproportionate increase in computational expense.

\subsection{Improvements}\label{sec:improvements}

Although the two hot spot geometries yield broadly consistent mass-radius constraints, \STPDT is, by construction, a restricted submodel of \PDTU. Together with the non-perfect posterior overlap and the substantially higher Bayesian evidence for \PDTU, this suggests that the \STPDT\ configuration may not fully capture the complexity of the surface emission. This is because the \STPDT\ model cannot mimic a temperature gradient in the southern-hemisphere hot spot, as it is restricted to a single uniform temperature circular region. The presence of two emission sub-regions for the secondary component in both spot models, and of an additional sub-region in the primary hot spot of \PDTU, hints at the existence of such a gradient. By contrast, the \PDTU\ parameterization can effectively mimic intra-spot temperature structure, naturally leading to an improved fit and higher evidence. This raises the possibility that a model consisting of only two hot spots, each with a smooth temperature gradient, might yield an even better description of the data than the current \PDTU\ setup, and that adding further geometric complexity would not necessarily improve the fit. Exploring this scenario, however, is not currently feasible within the existing \texttt{X-PSI} parameterization and would require substantial methodological development to implement and test.

We also note that \citet{Salmi:2023} found that, unlike for \joh, the assumed atmospheric composition and ionization state (i.e. whether partial ionization is allowed) for \jdbl\ can have a non-negligible impact on the inferred posterior distributions. In this work, we have restricted ourselves to a fully ionized hydrogen atmosphere, following previous analyses \citep{Miller:2019, Riley:2019, Vinciguerra:2024}, and did not explore alternative compositions due to computational limitations. Assessing the impact of atmospheric composition on the inferred mass--radius constraints and hot spot geometry for \jdbl\ is therefore left for future work.

Another important open question is whether the hot spot geometries inferred with the \STPDT and \PDTU models are physically realistic, given that some of the emitting regions have very small angular radii \citep[see Section~4 of][for an in-depth discussion]{Hoogkamer:2025}. It is also unclear whether these geometries are consistent with radio and gamma-ray observations \citep[see, e.g.,][]{Abdo:2013}. A joint radio–X-ray–gamma-ray analysis could therefore be used to discriminate between, or potentially rule out, one or both of these configurations \citep[see e.g.,][]{Bilous:2019, Chen:2020, Petri:2023, Cao:2026}. Such a multiwavelength approach can also provide complementary constraints on the viewing and magnetic geometry, which in turn would help reduce degeneracies in the pulse-profile modeling and strengthen the resulting mass–radius constraints. Exploring these multiwavelength constraints is beyond the scope of this study, but represents a promising direction for future work.

\section{Conclusion}\label{sec:conclusion}

We have presented an updated pulse profile modeling analysis of the MSP \jdbl\ using an expanded \NICER\ dataset accumulated over six years, jointly analyzed with archival \xmm\ observations to improve spectral background constraints. We modeled the phase and energy resolved emission with the two hot spot geometries identified by \citetalias{Vinciguerra:2024} as viable descriptions of the data: the \STPDT\ and \PDTU\ configurations. Both models provide a good description of the observed pulse profiles and yield broadly consistent constraints on the neutron-star mass and radius.

For the \STPDT\ model we infer $M = 1.29^{+0.13}_{-0.12}$ \msol and $R_{\rm eq} = 11.63^{+0.95}_{-0.89}$~km (68\% credible intervals). These constraints are essentially unchanged relative to \citetalias{Vinciguerra:2024}, indicating that the \STPDT\ mass--radius inference is stable despite the 50\% increase in \NICER\ counts, due to the improved sampler settings. For the \PDTU\ model we obtain $M = 1.43^{+0.20}_{-0.17}$ \msol\ and $R_{\rm eq} = 12.68^{+1.31}_{-1.04}$~km. Compared to the earlier \NICER\ dataset analyzed by \citetalias{Vinciguerra:2024}, the \PDTU\ posterior shifts substantially toward lower masses and radii, bringing its mass--radius constraints much closer to those obtained with \STPDT. At the same time, Bayesian model comparison still favors \PDTU, indicating that the data continue to support a more complex description of the surface emission.

The inferred hot spot properties from the two models are now broadly similar. In both cases, one hot spot is located in the northern hemisphere and exhibits a morphology suggestive of an intra-spot temperature gradient, characterized by a small, hot component embedded within a larger, cooler region. The \PDTU parameterization further allows for an additional small, hot component nested within a more extended and cooler region, a feature not available in \STPDT. If such a temperature gradient is indeed present on the stellar surface, this could naturally account for the higher Bayesian evidence obtained with the \PDTU model. However, the absence of prominent features in the residuals indicates that both hot spot models are able to reproduce the data adequately, with the \PDTU\ model yielding slightly more conservative constraints.

Further progress will require higher-quality data, as well as the exploration of physically motivated extensions to the emission model, such as hot spot prescriptions with smooth temperature gradients, and a systematic assessment of the impact of alternative atmospheric compositions, which have been shown to influence the inferred parameters for some MSPs. More broadly, the updated constraints reported here add to the growing set of \NICER pulse-profile measurements used to inform the neutron–star mass–radius relation and, ultimately, the dense-matter equation of state.

\section*{Acknowledgments}
Y.K., A.L.W. and M.H. acknowledge support  from NWO grant ENW-XL OCENW.XL21.XL21.038 \textit{Probing the phase diagram of Quantum Chromodynamics} (PI: Watts). T.S. acknowledges funding by the Research Council of Finland grant No.~368807. S.G., L.M., C.K., D.G.C., P.S. are supported by the CNES, and by the ANR (Agence Nationale de la Recherche) grant number ANR-20-CE31-0010 (ANR MORPHER). B.D. and A.L.W. acknowledge support from European Research Council (ERC) Consolidator Grant (CoG) No. 865768 AEONS (PI: Watts).  We acknowledge NWO for providing access to Snellius, hosted by SURF through the Computing Time on National Computer Facilities call for proposals. Some of the exploratory runs were carried out on the HELIOS cluster only on dedicated nodes funded through the ERC Consolidator Grant. Portions of this work carried out at NRL were funded by NASA.

\facilities{\NICER, XMM}

\software{Cython \citep{Behnel2011}, fgivenx \citep{Handley2018}, GetDist \citep{Lewis2019}, GNU Scientific Library (GSL; \citealt{Galassi2009}), HEASoft \citep{heasoft2014}, Matplotlib \citep{Hunter2007}, MPI for Python \citep{Dalcin2008}, \MultiNest \citep{multinest09}, nestcheck \citep{Higson2018JOSS}, NumPy \citep{Walt2011},\PyMultiNest \citep{PyMultiNest}, Python/C language \citep{Oliphant2007}, SciPy \citep{Jones}, \XPSI \citep{Riley:2023}.}

\bibliographystyle{aasjournal}
\bibliography{allbib}

\clearpage 
\appendix

\section{\texttt{Exploratory} runs}\label{appendix:runs}
\begin{figure*}
    \centering
     \includegraphics[width=0.45\linewidth]{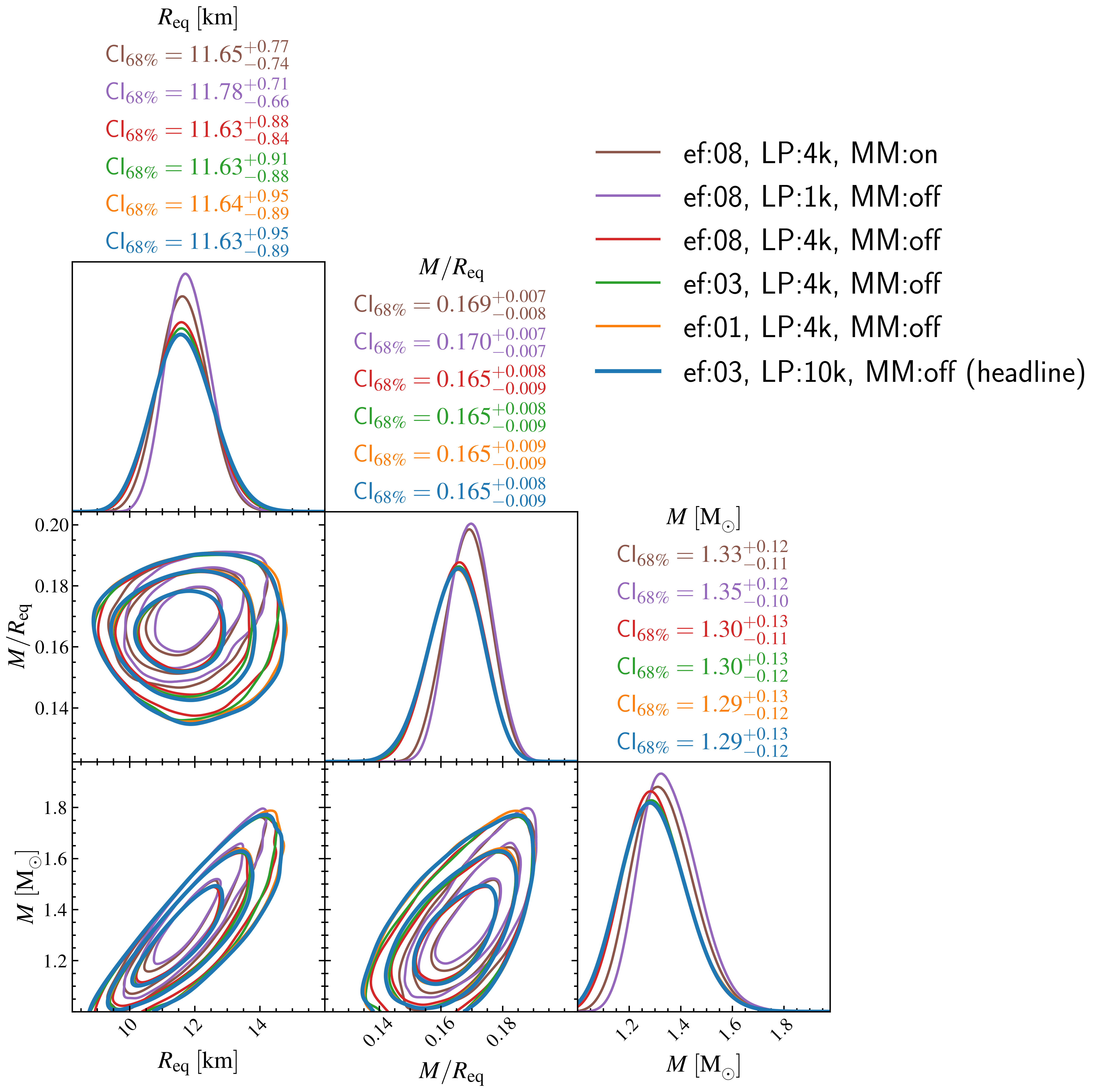}
     \includegraphics[width=0.45\linewidth]{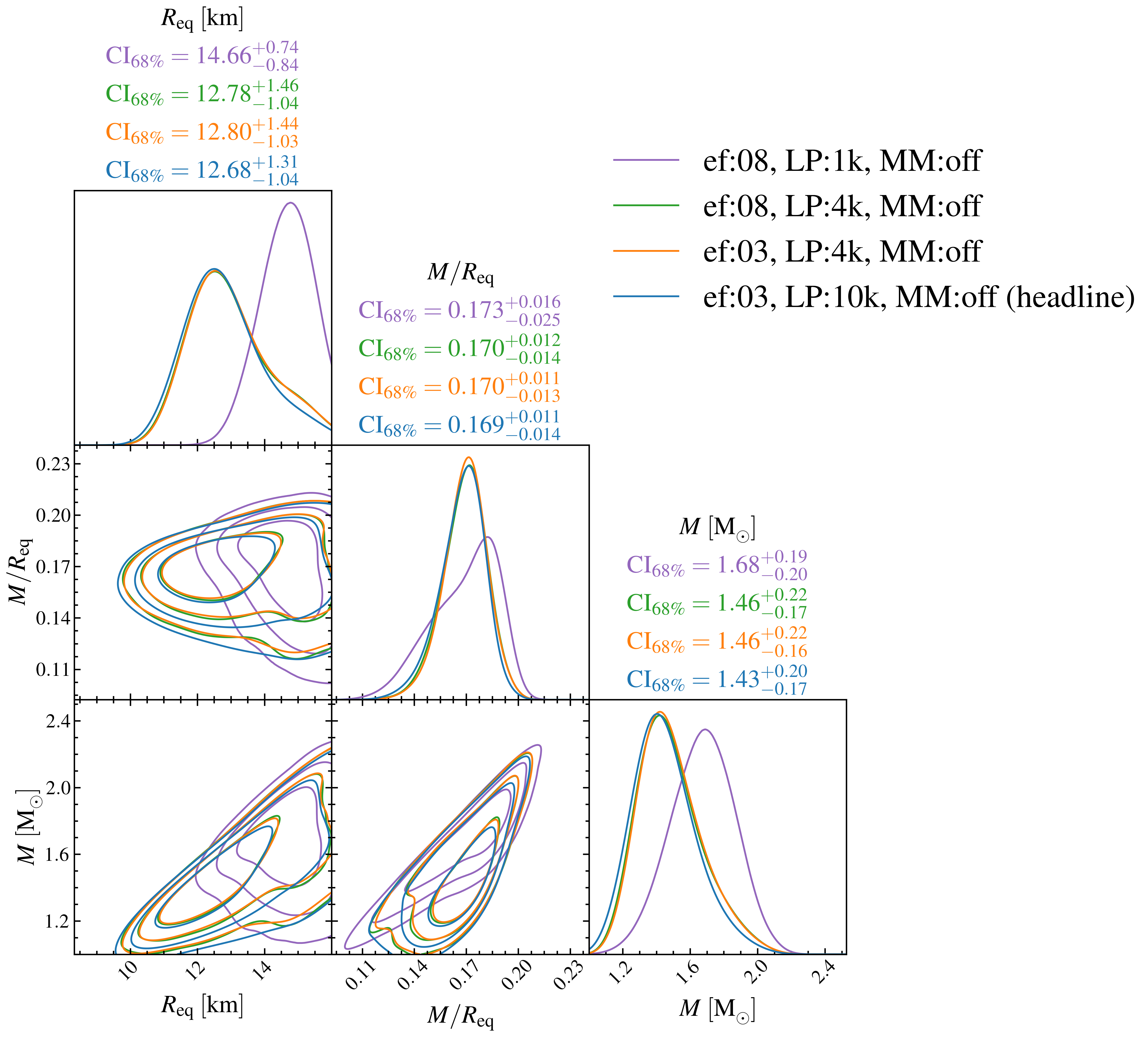}
      \caption{Posterior distributions of mass $M$, equatorial radius $R_{\rm eq}$, and compactness $M/R_{\rm eq}$ for PSR~J0030+0451 obtained using different sampler configurations to assess convergence. The left panel shows results for the \STPDT model, while the right panel corresponds to the \PDTU model. Labels of the form ef:xx, LP:yk, MM:zz denote the sampling efficiency (xx), the number of live points (y in thousands), and whether multimodal sampling is enabled (zz = on or off), respectively.}
  \label{fig:mass_radius_explore}
\end{figure*}

To assess convergence and the robustness of our results, we performed a series of exploratory runs to examine the sensitivity of the inferred parameters to the sampler settings (see Fig \ref{fig:mass_radius_explore}). In the \STPDT case, varying the sampling efficiency while keeping the other hyperparameters fixed yields nearly identical posterior distributions. The width of the mass credible interval mainly remains unchanged ($\Delta M \simeq 0.25\,M_\odot$), while the posterior median shifts by only $0.01\,M_\odot$ when the sampling efficiency is reduced to 0.1. For the radius, the median shifts by only $0.01\,\mathrm{km}$ over the same range. The width of the radius credible interval increases modestly, from 1.72 to 1.79 and 1.84\,km as the sampling efficiency is decreased from 0.8 to 0.3 and 0.1, respectively. This increase is marginal and is unlikely to have any significant impact on the inferred equation-of-state constraints. Similar trends are observed when varying the number of live points and for the \PDTU model.

\section{\texttt{Residuals}}\label{appendix:residuals}

\begin{figure*}
    \centering
    \includegraphics[width=0.45\linewidth]{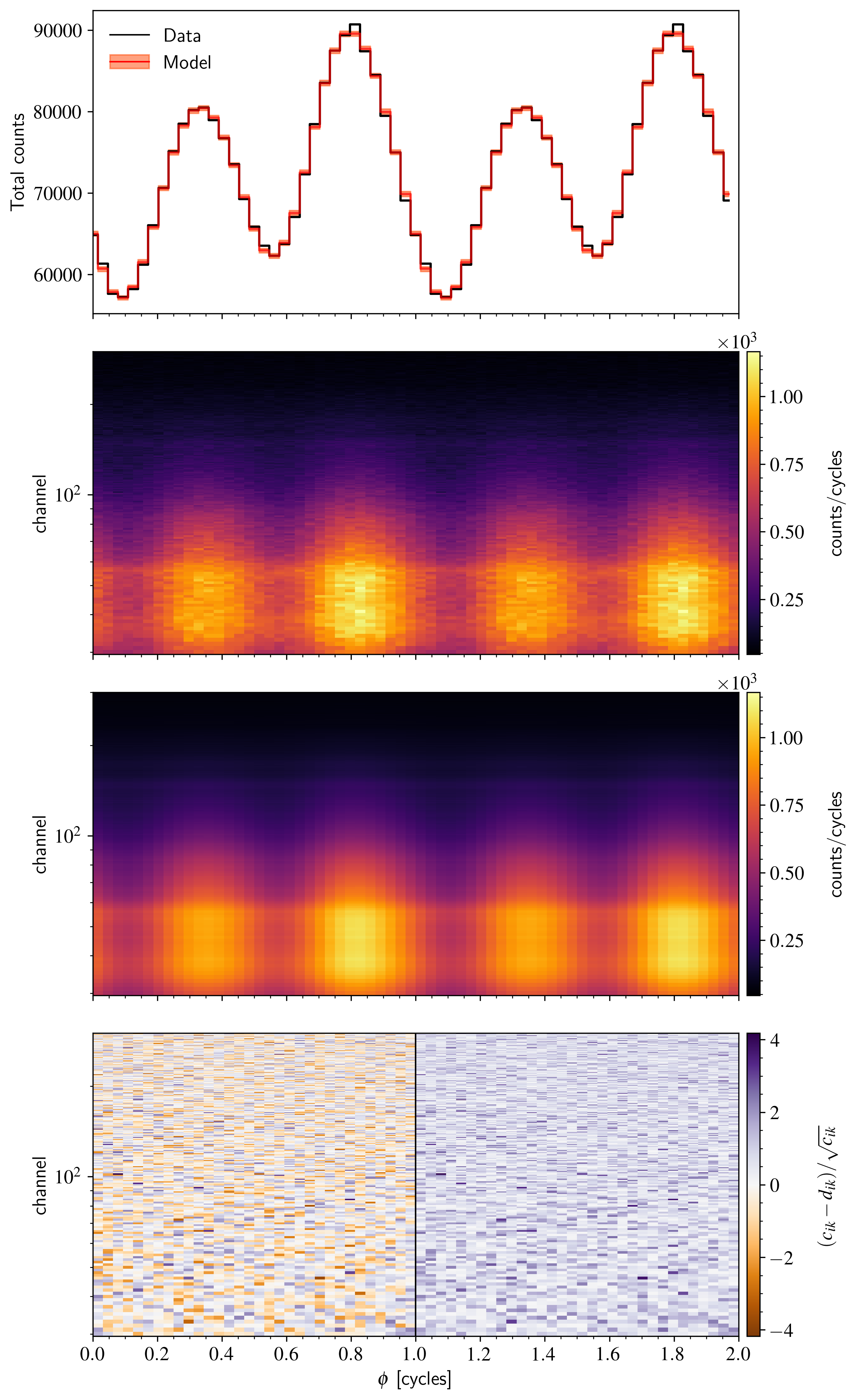}
    \includegraphics[width=0.45\linewidth]{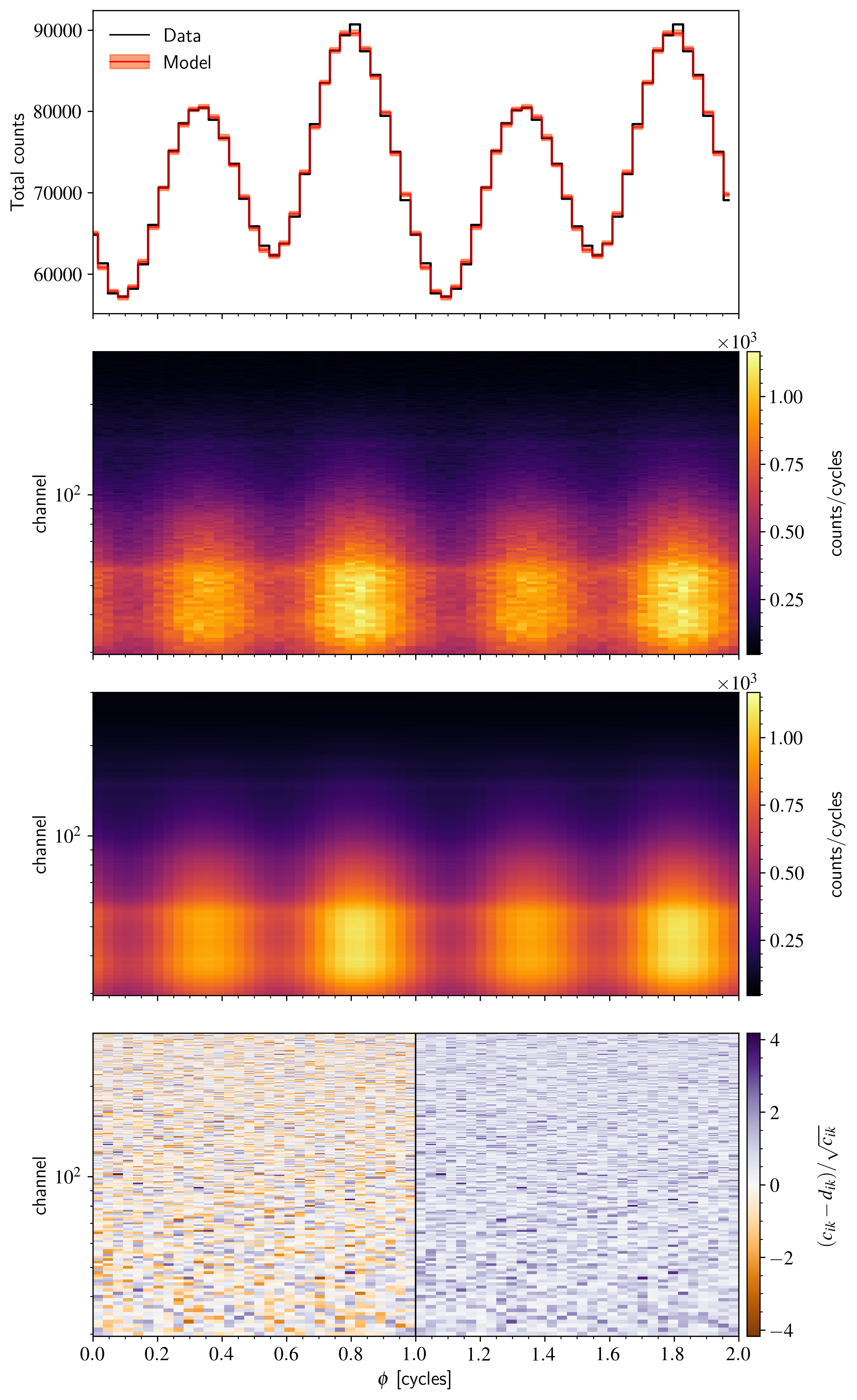}
    \caption{Data--model comparison and residuals for NICER. Each column corresponds to a different model: the left column shows the \STPDT model and the right column the \PDTU model. In each column, the top panel displays the observed pulse profile (data) summed over energy channels (black) together with the corresponding model pulse profile (orange), with the shaded region indicating the 68\% credible interval. The second panel shows the observed pulse profile (as in Fig.~\ref{fig:lightcures}). The  third panel shows  shows the model-averaged pulse profile computed from 200 posterior samples, and the bottom panel shows the corresponding residuals. Here $d_{ik}$ and $c_{ik}$ denote the data and model counts, respectively, in the $i$th rotational phase bin and $k$th energy channel. In the bottom panels, the left subpanel shows the signed residuals $(c_{ik}-d_{ik})/\sqrt{c_{ik}}$, while the right subpanel shows their absolute values.}
    \label{fig:residuals}
\end{figure*}

Figure~\ref{fig:residuals} shows the data--model comparison and residuals for NICER, evaluated using 200 random samples from the joint \NICER+\xmm posterior distribution.
The residuals show no evidence for phase or energy-dependent systematic trends indicative of model inadequacies. Both the \STPDT and \PDTU models provide comparably good descriptions of the new NICER data, with residuals that are visually similar across phase and energy.

\section{\texttt{Hot spot geometry}}\label{appendix:geometry}

\begin{figure*}
    \centering

    \begin{minipage}[t]{0.4\textwidth}
        \centering
        \includegraphics[width=\linewidth]{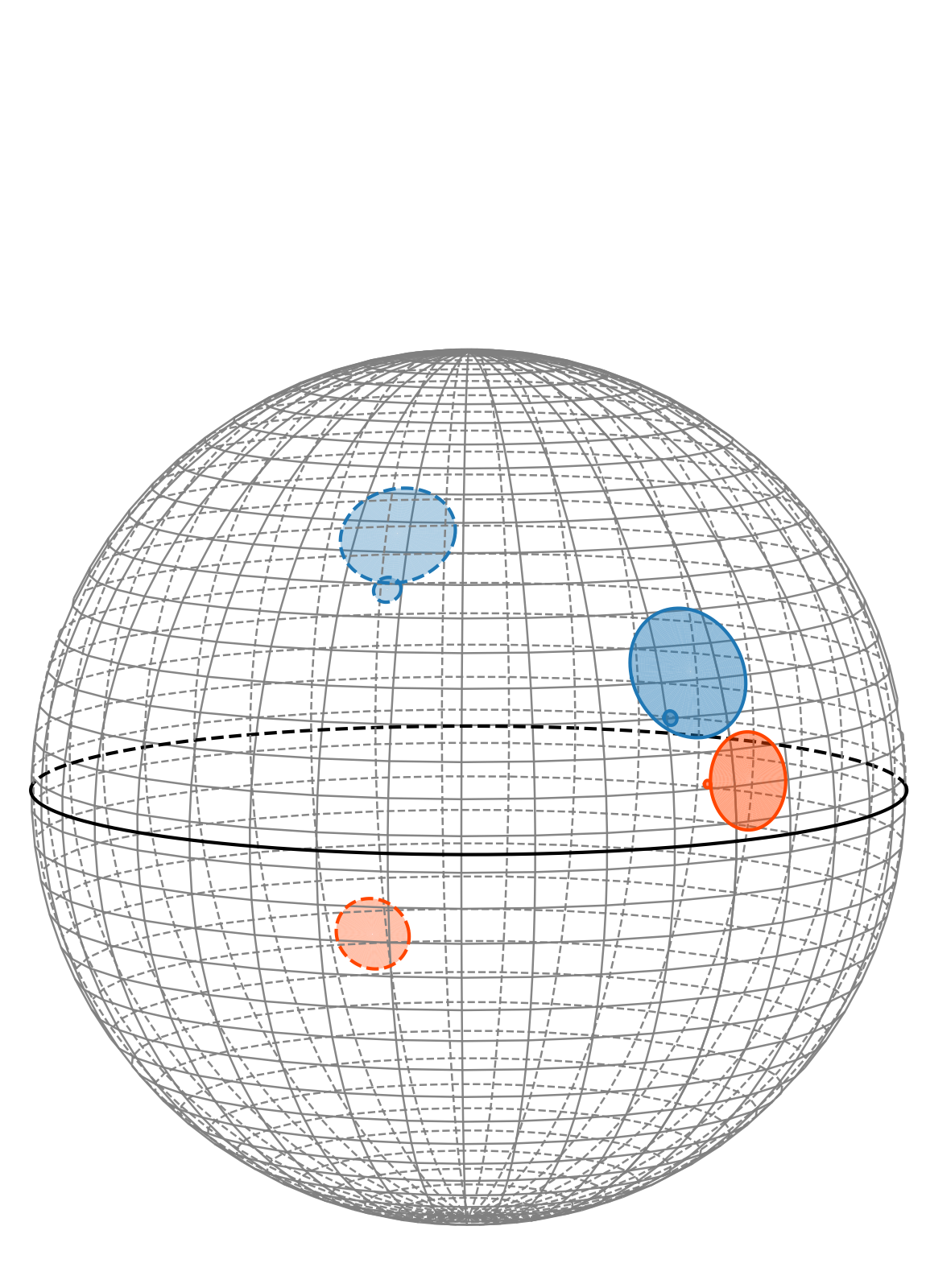}
    \end{minipage}\hspace{1cm}
    \begin{minipage}[t]{0.4\textwidth}
        \centering
        \includegraphics[width=\linewidth]{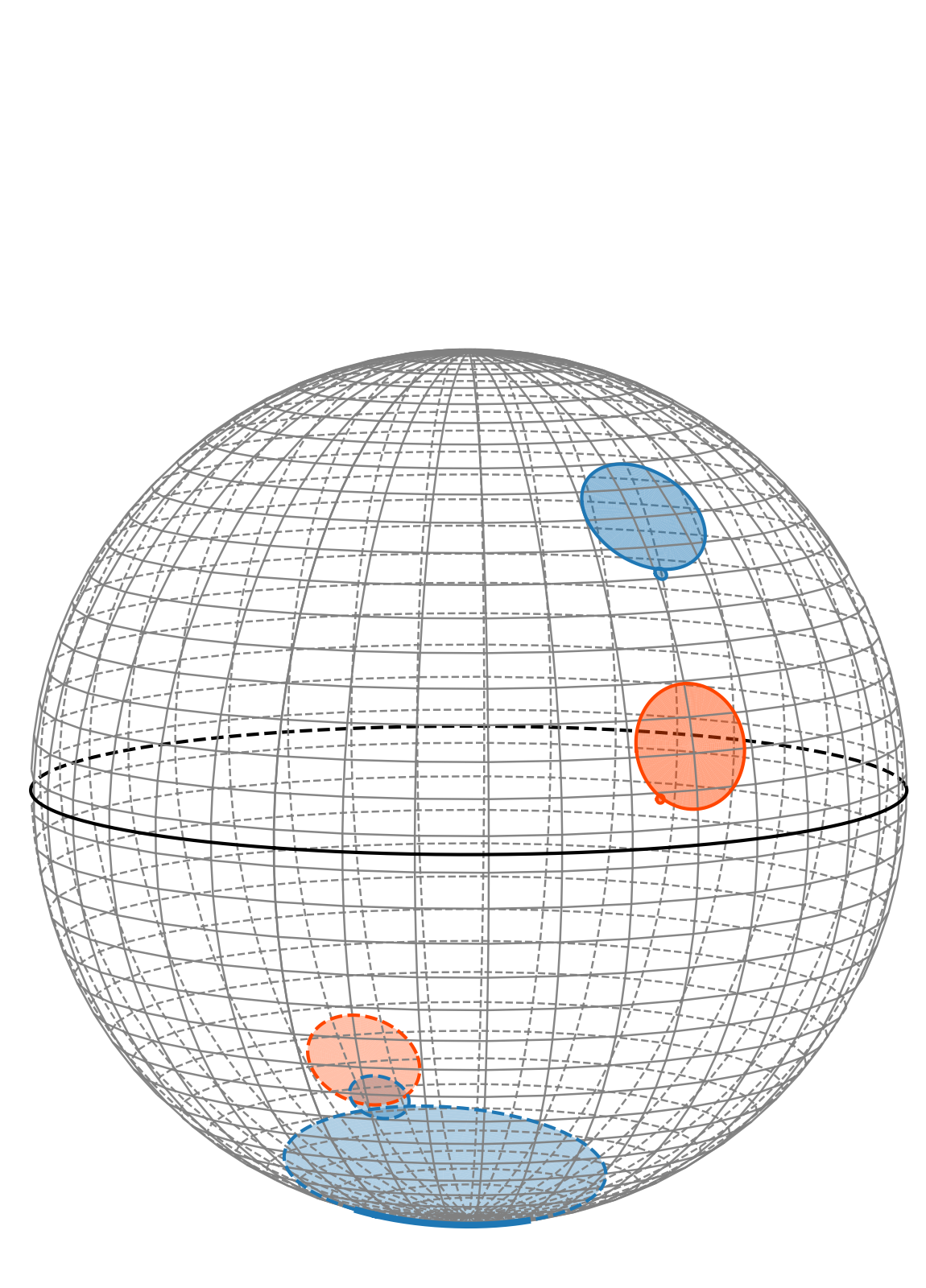}
    \end{minipage}

    \vspace{-2.cm}

    \begin{minipage}{\textwidth}
        \hspace{8cm}
        \includegraphics[width=0.15\textwidth]{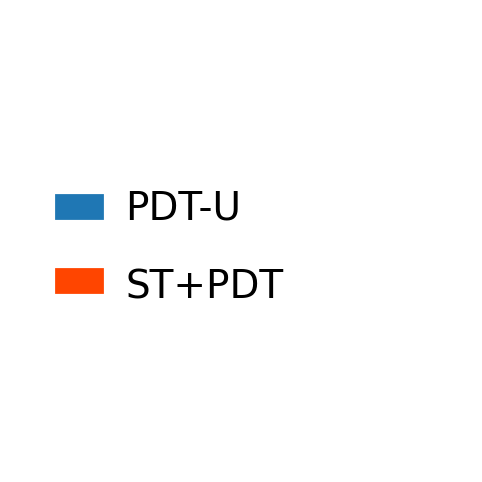}
    \end{minipage}

    \caption{Surface maps of \jdbl in the source frame for the maximum-likelihood parameter vectors. Left: configurations found by \citetalias{Vinciguerra:2024} for the ST+PDT (orange) and PDT-U (blue) components. Right: same as left but using the new dataset presented in this work. The black great circle marks the stellar equator. The maps are shown for an observer inclination of $i \approx 82^\circ$ which corresponds to the maximum-likelihood solution of the \PDTU model (the maximum-likelihood solution of the for ST+PDT is $i \approx 74^\circ$).}
    \label{fig:surf_maps}
\end{figure*}

In Figure~\ref{fig:surf_maps}, we present the surface maps of \jdbl in the source frame corresponding to the maximum-likelihood parameter vectors. The surface map obtained by \citetalias{Vinciguerra:2024} (left panel), assuming the \STPDT configuration, differs markedly from that found in the \PDTU case. In the \PDTU scenario, both hot spots are located in the same hemisphere. By contrast, in this work (right panel), we find that the two hot regions lie in opposite hemispheres for the hot-spot models with some similarities.

\section{\texttt{Summary Table for the \texttt{PDT-U} model}}\label{appendix:table_summary}

Table~\ref{table:summary} summarizes the results for the \PDTU\ model. We retain the same hotspot nomenclature as in previous analyses (see, e.g., \citetalias{Vinciguerra:2024}, \citealt{Salmi:2024_1231}) to enable direct comparison, ensure reproducibility, and facilitate its use in subsequent modelling.

\begin{deluxetable*}{lccccc}
\caption{Summary Table for the \texttt{PDT-U} model}\label{table:summary}
\tablehead{
\colhead{Parameter} & \colhead{Description} & \colhead{Prior PDF (Density and Support)} & \colhead{$\widehat{\textrm{CI}}_{68\%}$} & \colhead{$\widehat{D}_{\textrm{KL}}$} & \colhead{$\widehat{\textrm{ML}}$}
}
\startdata
$P$ $[$ms$]$ & Coordinate spin period & $P=4.87$, fixed & $-$ & $-$ & $-$\\
\hline
$M$ $[M_{\odot}]$ & Gravitational mass & $M\sim\mathcal{U}(1.0,3.0)$ & $1.43_{-0.17}^{+0.20}$ & $0.96$ & $1.18$ \\ 
$\cos(i)$ & Cosine Earth inclination to spin axis & $\cos(i)\sim\mathcal{U}(0.0,1.0)$  &$0.279_{-0.075}^{+0.076}$ & $1.62$ & $0.147$ \\ 
\hline
$R_{\textrm{eq}}$ $[$km$]$ &
Coordinate equatorial radius &$R_{\rm eq}\sim\mathcal{U}(3r_g(1.0),16.0)^{\mathrm{a}}$&$12.68_{-1.04}^{+1.31}$ & $0.74$ & $14.09$ \\ 
&With compactness condition & $R_{\textrm{polar}}/r_{\rm g}(M)>3$\\
&With surface gravity condition & $13.7\leq \log_{10}g(\theta)\leq15.0$,~$\forall\theta$\\
\hline
$\Theta_{p}$ $[$radians$]$ & $p$ superseding component center colatitude & $\cos(\Theta_{p})\sim U(-1,1)$ & $1.33_{-0.16}^{+0.18}$ & $1.23$ & $0.95$ \\ 
$\Theta_{c,p}$ $[$radians$]$ &
$p$ ceding component center colatitude &$\cos(\Theta_{c,p})\sim U(-1,1)$ &$1.31_{-0.17}^{+0.17}$ &$1.25$ &$0.796$ \\ 
$\Theta_{s}$ $[$radians$]$ & $s$ superseding component center colatitude & $\cos(\Theta_{s})\sim U(-1,1)$ & $2.34_{-0.18}^{+0.08}$ & $1.44$ & $2.532$ \\ 
$\Theta_{c,s}$ $[$radians$]$ &
$s$ ceding component center colatitude & $\cos(\Theta_{c,s})\sim U(-1,1)$ & $2.77_{-0.63}^{+0.14}$ & $1.16$ & $2.95$ \\ 
$\phi_{p}$ $[$cycles$]$ & $p$ superseding component initial phase & $\phi_{p}\sim U(-0.25,0.75)$, wrapped & $0.152_{-0.008}^{+0.005}$ & $2.69$ & $0.156$\\ 
$\phi_{s}$ $[$cycles$]$ &$s$ superseding component initial phase &$\phi_{s}\sim U(-0.25,0.75)$, wrapped &$0.127_{-0.008}^{+0.004}$ & $2.84$ &$0.130$ \\ 
$\chi_{p}$ $[$radians$]$ & Azimuthal offset between the $p$ components & $\chi_{p}\sim U(-\pi,\pi)$ & $-0.041_{-0.024}^{+0.031}$ &$3.02$ &$-0.035$\\ 
$\chi_{s}$ $[$radians$]$ &
Azimuthal offset between the $s$ components & $\chi_{s}\sim U(-\pi,\pi)$ &$-0.004_{-0.0497}^{+0.0503}$ & $2.20$ &$0.063$\\ 
$\zeta_{p}$ $[$radians$]$ &$p$ superseding component angular radius & $\zeta_{p}\sim U(0,\pi/2)$ & $0.017_{-0.003}^{+0.006}$ & $4.20$ & $0.0141$ \\ 
$\zeta_{c,p}$ $[$radians$]$ &
$p$ ceding component angular radius &
$\zeta_{c,p}\sim U(0,\pi/2)$ &
$0.150_{-0.018}^{+0.023}$ &
$2.8$ &
$0.15$ \\ 
$\zeta_{s}$ $[$radians$]$ &
$s$ superseding region angular radius &
$\zeta_{s}\sim U(0,\pi/2)$ &
$0.0810_{-0.015}^{+0.011}$ &
$3.21$ &
$0.07$ \\ 
$\zeta_{c,s}$ $[$radians$]$ &
$s$ ceding component angular radius &
$\zeta_{s,p}\sim U(0,\pi/2)$ &
$0.59_{-0.22}^{+0.60}$ &
$0.54$ &
$0.38$ \\ 
&No region-exchange degeneracy & $\Theta_{s}\geq\Theta_{p}$\\
&Nonoverlapping hot regions & function of all $\Theta$, $\phi$, $\chi$, and $\zeta$\\
&Overlapping hot region components$^{\mathrm{b}}$ & function of all $\Theta$, $\phi$, $\chi$, and $\zeta$\\
\hline
$\log_{10}\left(T_{p}\;[\textrm{K}]\right)$ &
$p$ superseding component effective temperature  &
$\log_{10}\left(T_{p}\right)\sim U(5.1,6.8)$, \TT{NSX} limits &
$6.28_{-0.04}^{+0.03}$ &
$3.3591$ &
$6.289$ \\ 
$\log_{10}\left(T_{c,p}\;[\textrm{K}]\right)$ &
$p$ ceding component effective temperature  &
$\log_{10}\left(T_{c,p}\right)\sim U(5.1,6.8)$, \TT{NSX} limits &
$5.89_{-0.04}^{+0.03}$ &
$3.34$ &
$5.88$ \\ 
$\log_{10}\left(T_{s}\;[\textrm{K}]\right)$ &
$s$ superseding component effective temperature  &
$\log_{10}\left(T_{s}\right)\sim U(5.1,6.8)$, \TT{NSX} limits &
$6.10_{-0.01}^{+0.02}$ &
$4.13$ &
$6.10$ \\ 
$\log_{10}\left(T_{c,s}\;[\textrm{K}]\right)$ &
$s$ ceding component effective temperature  &
$\log_{10}\left(T_{c,s}\right)\sim U(5.1,6.8)$, \TT{NSX} limits &
$5.67_{-0.24}^{+0.12}$ &
$1.61$ &
$5.79$ \\ 
$D$ $[$kpc$]$ &
Earth distance &
$D\sim\mathcal{N}(0.325,0.009^2)$\ &
$0.327_{-0.008}^{+0.007}$ &
$0.06$ &
$0.325$ \\ 
$N_{\textrm{H}}$ $[10^{20}$cm$^{-2}]$ &
Interstellar neutral H column density &
$N_{\textrm{H}}\sim U(0.0,5.0)$ &
$1.17_{-0.23}^{+0.25}$ &
$2.30$ &
$1.17$ \\ 
\hline
$\alpha_{\rm XTI}$ & NICER effective-area scale factor &$\alpha_{\rm XTI}\sim\mathcal{N}(1.0,0.104^2)$& $0.920_{-0.079}^{+0.068}$ & $0.52$ & $0.827$ \\ 
$\alpha_{\rm MOS1}$ & XMM–Newton MOS1 effective-area scale factor &$\begin{aligned} \alpha_{\rm MOS1}\mid\alpha_{\rm XTI} \sim
\mathcal{N}\big(&1+\rho(\alpha_{\rm XTI}-1),\\ &0.104^2(1-\rho^2)\big)^{\mathrm{b}}\end{aligned}$\\
\hline
\hline
&Sampling Process Information&&& \\
\hline
&Number of free parameters: $23$ &&& \\
&Number of live points: $10{,}000$ &&& \\
&Sampling efficiency (SE): $0.3$ &&& \\
&Termination condition: $0.1$ &&& \\
&Evidence: $\widehat{\ln\mathcal{Z}}= -44475.77\pm0.08$ &&&\\ 
&Number of core hours: $4{,}802{,}880$ &&& \\
&Likelihood evaluations: $163{,}156{,}336$ &&& \\
\enddata
\tablecomments{\ \
We show the prior PDFs, $68.3\,\%$ credible intervals around the median $\widehat{\textrm{CI}}_{68\%}$, KL-divergence $\widehat{D}_{\textrm{KL}}$ in \textit{bits}, and the maximum-likelihood nested sample $\widehat{\textrm{ML}}$ for all the parameters. 
The subscripts $p$ and $s$ denote for primary and secondary hot region parameters, respectively. Note that $\phi_p$ ($\phi_s$) is measured with respect to the meridian on which Earth (Earth antipode) lies.
\\
$^{\mathrm{a}}$ Truncated between $3r_{\rm g}(1) \approx 4.4$ km and $16$ km. \\
$^{\mathrm{b}}$ The NICER and XMM--Newton effective-area scale factors are treated as correlated to account for shared calibration uncertainties; the conditional form is used to implement this correlation with $\rho=0.916$.\\
$^{\mathrm{c}}$ Ensuring that superseding components never engulf the corresponding ceding components.
\\
}
\end{deluxetable*}

\end{document}